\documentclass[twocolumn]{aastex631}
\usepackage{natbib}
\usepackage{amsmath}
\usepackage{bm}
\usepackage{graphicx}
\usepackage{multirow}
\usepackage{booktabs} 
\usepackage{array}
\usepackage{xcolor} 
\usepackage{hyperref}
\usepackage{soul}

\begin{document}

\title{Stereoscopic Observations of Solar X-ray Sources Explained by a Data-Constrained Magnetohydrodynamic Simulation}

\author[0000-0003-2002-0247]{Keitarou Matsumoto}
\affiliation{Center for Solar-Terrestrial Research, New Jersey Institute of Technology, Newark, NJ 07102-1982, USA}

\author[0000-0001-5121-5122]{Satoshi Inoue}
\affiliation{Center for Solar-Terrestrial Research, New Jersey Institute of Technology, Newark, NJ 07102-1982, USA}

\author[0000-0002-2633-3562
]{Meiqi Wang}
\affiliation{Center for Solar-Terrestrial Research, New Jersey Institute of Technology, Newark, NJ 07102-1982, USA}

\author[0000-0002-2002-9180]{Säm Krucker}
\affiliation{University of Applied Sciences and Arts Northwestern Switzerland, Bahnhofstrasse 6, 5210 Windisch, Switzerland}
\affiliation{Space Sciences Laboratory, University of California, 7 Gauss Way, 94720 Berkeley, USA}

\author[0000-0001-5037-9758
]{Satoshi Masuda}
\affiliation{Institute for Space-Earth Environmental Research, Nagoya University, Furo-cho, Chikusa-ku, Nagoya,
Aichi 464-8601, Japan}

\author[0000-0002-8538-3455]{Muriel Zoë Stiefel}
\affiliation{University of Applied Sciences and Arts Northwestern Switzerland, Bahnhofstrasse 6, 5210 Windisch, Switzerland}
\affiliation{ETH Zürich, Rämistrasse 101, 8092 Zürich, Switzerland}

\author[0000-0002-5865-7924]{Jeongwoo Lee}
\affiliation{Center for Solar-Terrestrial Research, New Jersey Institute of Technology, Newark, NJ 07102-1982, USA}

\author[0000-0002-0660-3350]{Bin Chen}
\affiliation{Center for Solar-Terrestrial Research, New Jersey Institute of Technology, Newark, NJ 07102-1982, USA}

\author[0000-0002-5233-565X]{Haimin Wang}
\affiliation{Center for Solar-Terrestrial Research, New Jersey Institute of Technology, Newark, NJ 07102-1982, USA}

\email{km876@njit.edu}



\begin{abstract}

We investigated the three-dimensional (3D) magnetic structures and dynamics responsible for particle acceleration in an X7.1-class flare that occurred on October 1, 2024, in NOAA active region 13842. We combined stereoscopic hard X-ray (HXR) observations from the Advanced Space-based Solar Observatory/Hard X-ray Imager (HXI) and the Solar Orbiter/Spectrometer Telescope for Imaging X-rays (STIX) with a 3D magnetohydrodynamic (MHD) simulation constrained by observed photospheric magnetic fields. During the two main peaks of the impulsive phase, HXR footpoints appeared at different locations, indicating a migration of the primary reconnection site in the corona. Our data-constrained MHD simulation successfully reproduced the reconnected field lines linking the observed conjugate HXR footpoints. Furthermore, the simulation shows that these primary reconnections occur along a single quasi-separatrix layer (QSL) system. Therefore, the two main peaks of HXR can be interpreted as episodic energy release within the single QSL system. This study demonstrates that the data-constrained MHD model provides a realistic 3D magnetic context for interpreting HXR emission. Notably, STIX observations revealed a vertically distributed thermal HXR source, extending from the footpoints to the looptop, with its centroid migrating between the two peaks. This marks a first step toward understanding the particle acceleration processes in solar flares.

\end{abstract}

\keywords{Solar flares (1496) --- Magnetohydrodynamics (1964) --- Solar active region
magnetic fields (1975) --- Magnetohydrodynamical simulations (1966) --- Non-thermal radiation sources (1119) --- Solar x-ray emission (1536)}


\section{Introduction} \label{sec:intro}
Solar flares release magnetic energy, and a portion of this energy goes into accelerating particles \citep{Emslie2012}. Accelerated electrons produce non-thermal hard X-rays (HXR) and microwaves, making particle acceleration central to both solar and plasma physics \citep{Benz2017}. Since the first report of the ``Masuda flare'' \citep{Masuda1994}, the above-the-loop-top (ALT) region has been considered a main candidate for the primary particle acceleration site \citep{Krucker2010, Krucker2014, Fleishman2022}. Recently, a “magnetic bottle’’ structure at the ALT region was revealed using microwave imaging spectroscopy observations made by the Expanded Owens Valley Solar Array (EOVSA; \citealt{Gary2018}). This region coincides with strong HXR/Microwave emission and concentrated high-energy electrons \citep{Chen2020,Chen2024}, indicating efficient acceleration and confinement. However, a detailed understanding of the particle acceleration and transport processes requires bridging the enormous scale gap between kinetic processes and macroscopic magnetohydrodynamics (MHD) dynamics. Recently, macroscopic models that capture the particle acceleration and transport processes in MHD-scale simulation domains have shed light on the role of reconnection-driven structures such as magnetic bottles, termination shocks, magnetic islands, plasma outflows, and turbulence in non-thermal particle production \citep{Kong2019, Arnold2021, Li2022, Li2025, Chen2024, Yin2024, Hu2025}.

Three-dimensional (3D) coronal fields are routinely reconstructed with nonlinear force-free field (NLFFF) extrapolations \citep{Wiegelmann2012, Inoue2016}. In low-$\beta$ corona (0.01 to 0.1; \citealt{Gary2001}), where $\beta$ is the ratio of gas pressure to magnetic pressure, magnetic fields dominate and the force-free approximation is reasonable, but NLFFF is static and cannot reproduce eruptive evolution. Data-constrained MHD simulations address this limitation by imposing observed photospheric fields as boundary conditions and have reproduced key observational features \citep{Inoue2015, Jiang2018, Yamasaki2022, Kumar2025, Liu2025, Matsumoto2025b}.

Stereoscopic X-ray observations offer a powerful probe of 3D particle acceleration. \citet{Ryan2024b} combined Solar Orbiter/Spectrometer Telescope for Imaging X-rays (STIX; \citealt{Krucker2020}) with Hinode/X-Ray Telescope (XRT; \citealt{Golub2007}), and \citet{Ryan2024a} paired STIX with Advanced Space-based Solar Observatory/Hard X-ray Imager (ASO-S/HXI; \citealt{Zhang2019}). Both studies aim to reconstruct the 3D magnetic geometry of X-ray sources and, ultimately, the full 3D X-ray structure. In this study, we analyze the X7.1 flare using extreme ultraviolet (EUV) imaging data from Solar Dynamics Observatory/Atmospheric Imaging Assembly (SDO/AIA; \citealt{Lemen2012}) and photospheric magnetic field measurements from SDO/Helioseismic and Magnetic Imager (HMI; \citealt{Scherrer2012}), together with X-ray observations made by HXI and STIX from two different viewing perspectives, to interpret the event stereoscopically in combination with data-constrained MHD simulations. Our goal is to identify the 3D magnetic structures responsible for the observed acceleration and to lay the groundwork for quantitative constraints.

To test inferences about energetic electrons against observation, it's necessary to synthesize the HXR or microwave emission produced in a given magnetic geometry. \citet{Gordovskyy2020} reproduced the spatial locations of observed HXR footpoints by combining data-constrained MHD simulations with a calculated map of HXR emission through test-particle calculations. However, the validation of the temporal evolution and 3D structure of the HXR sources remained to be addressed. A systematic framework is provided by \texttt{GX Simulator} \citep{Nita2023ApJS}, which populates a 3D magnetic model with thermal plasma and non-thermal electrons and computes HXR/Microwave images and spectra for direct comparison with observations. This enables consistency tests of the modeled fields and non-thermal emission \citep{Kuroda2018, Kuroda2020, Fleishman2021, Fleishman2023, Fleishman2025}.

In this paper, we study stereoscopic X-ray observations of a solar flare from NOAA AR 13842 in comparison with a data-constrained MHD simulation. NOAA AR 13842 produced two major events, which are an X7.1 flare on 2024-10-01 (SOL2024-10-01T22:20) and an X9.0 flare on 2024-10-03 (SOL2024-10-03T12:18). Our MHD simulations (\citealt{Matsumoto2025a, Matsumoto2025b}) reproduced their eruptions, which resulted from torus instability (\citealt{Bateman1978, Kliem2006}) and reconnection. Here, as a first step toward understanding non-thermal emission in realistic 3D fields, we compare HXR observations with data-constrained MHD evolution of the eruption. Section \ref{sec:sec2} describes observations and simulation methods, Section \ref{sec:sec3} presents results, and Section \ref{sec:sec4} discusses implications. Conclusions appear in Section \ref{sec:sec5}.

\section{Observations and MHD simulations} \label{sec:sec2}
\subsection{Observation}\label{sec:obs}
\begin{figure*}[t]
\plotone{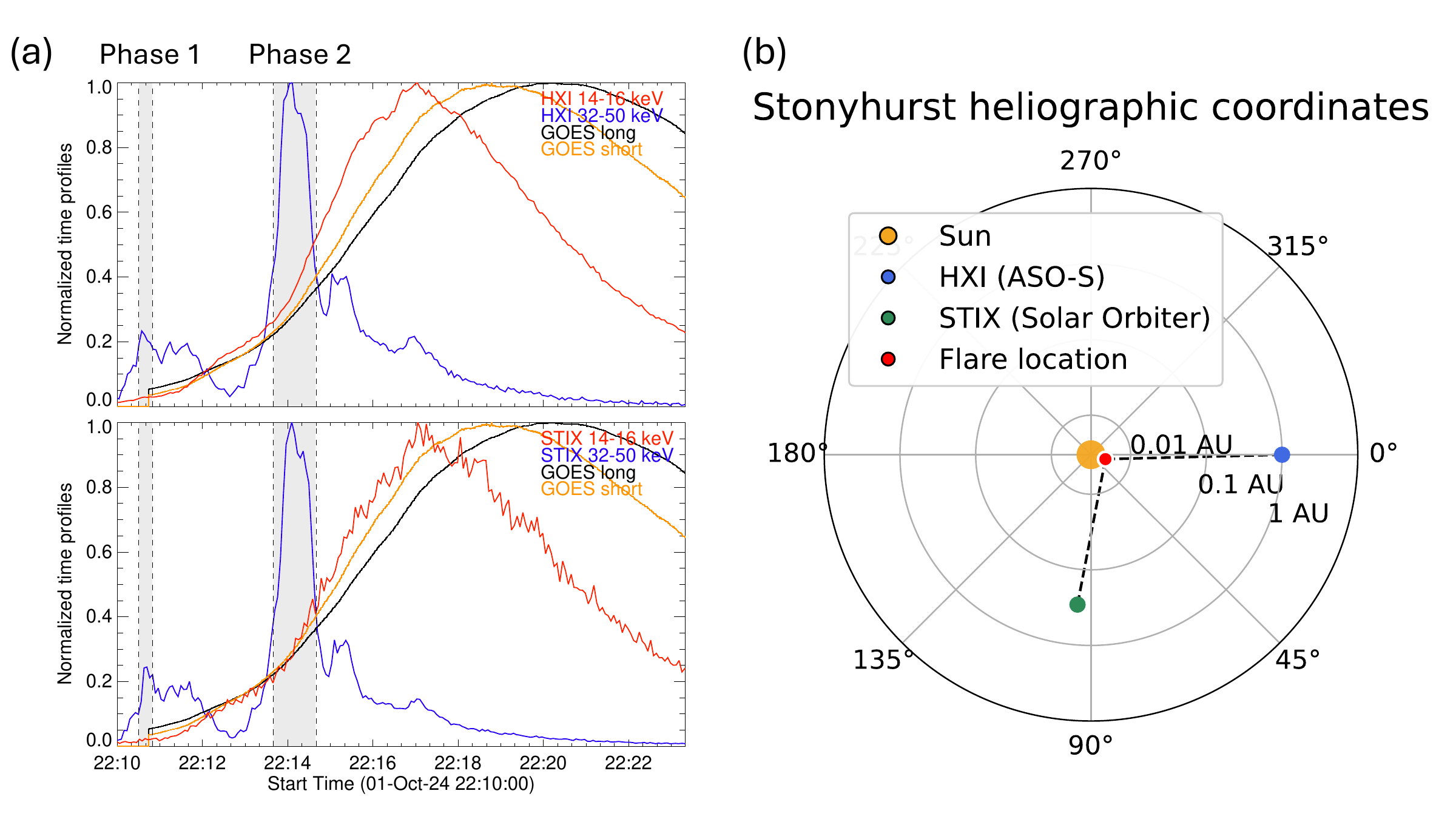}
\caption{(a) Temporal evolution of X-ray observed by HXI, STIX, and GOES-18. For both HXI and STIX, the data are summed to a 4 s resolution. The blue, red, light blue, and green curves denote 32--50 keV, 14-16 keV, 0.5-4.0 \AA, and 1.0-8.0 \AA, respectively. The STIX light curves are shifted by 352.9 s to consider the different travel time of the light. For 14-16 keV in STIX, the background detector with pixels 2 \& 5 was used. The two gray shaded regions indicate the time intervals 22:10:30--22:10:50 (Phase 1) and 22:13:40--22:14:40 (Phase 2), which are used for hard X-ray imaging. (b) Stonyhurst heliographic coordinates, showing the positions of HXI, STIX, and flare location.
}\label{fig:fig1}
\end{figure*}

The X7.1 flare occurred on 2024 October 1 in NOAA AR13842. It started at 21:58 UT and peaked at 22:20 UT in the Geostationary Operational Environmental Satellite (GOES) soft X-ray light curve. Figure \ref{fig:fig1}(a) shows the light curves observed by HXI, STIX, and GOES. The STIX time profile was constructed by summing over 8 pixels from the 24 coarsest imaging detectors, while the HXI time profile was obtained by summing the signals from the two thick detectors for the total flux (D92 and D93). In this study, since the light travel time is different between HXI and STIX, the observational time of STIX is shifted by the difference in travel time (352.9 s) to match the HXI observation. The attenuator was inserted in STIX around 22:04:42 UT, corresponding to 22:10:34 UT in Figure \ref{fig:fig1}(a). Therefore, the STIX background detector with pixels 2 \& 5 profile \citep{Stiefel2025} is used to get an unattenuated time profile at 14-16 keV for Figure \ref{fig:fig1}(a). The STIX background profile in 14-16 keV has some fluctuation because we used only two pixels, which are covered by the middle sized openings and suitable to analyze the large flares \citep{Stiefel2025}.

To compare the HXR observations with the MHD simulation results \citep{Matsumoto2025b}, we focused on the impulsive phase of the HXR emission, as indicated by the two time intervals (Phases 1 and 
 2) in Figure \ref{fig:fig1}(a). Figure \ref{fig:fig1}(b) shows the Stonyhurst heliographic coordinates, indicating the positions of HXI, STIX, and the flare site. From the Earth’s perspective (HXI view), the flare occurred at ($x, y$) = ($-$270 arcsec, $-$470 arcsec) relative to the disk center in the Helioprojective Cartesian coordinate system, whereas from the STIX perspective, it was located near the west limb. The separation angle between HXI and STIX was $\sim95^{\circ}$, providing a unique opportunity to interpret the X-ray emission from two distinct viewpoints, as shown in Figure \ref{fig:fig1}(b).

 However, a reliable comparison of the reconstructed images from HXI and STIX requires accurate co-alignment between the two instruments. Previous studies (e.g., \citealt{Massa2022}; \citealt{Ryan2024a}) have used the ultraviolet (UV) flare ribbons observed by AIA as a reference for shifting HXR sources in both HXI and STIX. Co-alignment criteria and applied shifts are described in Sections \ref{sec:sec3.1} and \ref{sec:sec3.3}. For image reconstruction, the CLEAN algorithm was applied to HXI, while STIX images were synthesized using the Maximum Entropy Method (MEM\_GE; \citealt{Massa2020}). As reported by \citet{Massa2022}, MEM\_GE generally achieves a smaller $\chi^{2}$ with respect to the observed visibilities compared to CLEAN, although both methods yield consistent large-scale HXR morphologies. In this study, we focus on identifying the locations of the HXR footpoints rather than discussing finer structures. Therefore, we use HXI subcollimators $\ge 13.4$ arcsec (D39–D91) and a 20 arcsec beam for STIX. The results of these imaging analyses are presented in Section \ref{sec:sec3.1} (HXI) and Section \ref{sec:sec3.3} (STIX).

\subsection{Nonlinear Force-free Extrapolation}\label{sec:nlfff}
For the NLFFF extrapolation, we used the SDO/HMI vector magnetogram at 20:36 UT (approximately one hour before the flare) in the cylindrical equal-area (CEA) projection as the bottom boundary. The NLFFF and MHD simulations followed  \citet{Matsumoto2025b}. To assist the reader, we provide a brief summary of the numerical approach. The governing equations of the NLFFF and MHD simulations are as follows:
\begin{equation}
\mathit{\rho} = \lvert \bm{B} \rvert,
\label{eq:eq1}
\end{equation}
\begin{equation}
\frac{\partial \bm{v}}{\partial t} = -(\bm{v} \cdot \nabla)\bm{v} + \frac{1}{\mathit{\rho}} \bm{J} \times \bm{B} + \mathit{\nu} \nabla^2 \bm{v},
\label{eq:eq2}
\end{equation}
\begin{equation}
\frac{\partial \bm{B}}{\partial t} = \nabla \times (\bm{v} \times \bm{B})+\eta \bm{\nabla}^2\bm{B} - \nabla \bm{\phi},
\label{eq:eq3}
\end{equation}
\begin{equation}
\bm{J} = \nabla \times \bm{B},
\label{eq:eq4}
\end{equation}
\begin{equation}
\frac{\partial \bm{\phi}}{\partial t} + c_h^2 \nabla \cdot \bm{B} = - \frac{c_h^2}{c_p^2} \bm{\phi},
\label{eq:eq5}
\end{equation}
where $\mathit{\rho}$, $\bm{B}$, $\bm{v}$, $\bm{J}$, and $\mathit{\phi}$ are plasma density, magnetic flux density, velocity, electric current density, and scalar potential, respectively. Length, magnetic field, plasma density, velocity, time, and 
electric current density are normalized by $L^{*}$ = 362.5 Mm, $B^{*}$ = $2.915\times 10^{-1}$ T, $\rho^*$ (kg m$^{-3})$, which corresponds to the density at the bottom surface of the simulation box, $V_A^* = {B^*}/{(\mu_0 \rho^*)^{1/2}}$ (m s$^{-1}$), $\tau_A^* = {L^*}/{V_A^*}$ (s), and $J^* = {B^*}/{\mu_0 L^*}$ (A m$^{-2}$). Here, 
$\mu_0$ is the magnetic permeability,

Plasma density, $\mathit{\rho}$, is assumed to be proportional to $\lvert \bm{B} \rvert$, reflecting that Alfvén waves propagate faster in regions of weaker magnetic field. The scalar potential $\phi$ is introduced to reduce errors in maintaining ${\bf \nabla }\cdot{\bm{B}}$ \citep{Dedner2002}. The coefficients $\nu$ and $\eta$ are viscosity and electric resistivity. In the NLFFF extrapolation, $\nu=1.0\times 10^{-3}$ and $\eta=5.0\times 10^{-5}+1.0\times 10^{-3}|{\bm{J}}\times {\bm{B}}||{\bm{v}}|^2/|{\bm{B}}|^2$. The second term of $\eta$ is included to accelerate the relaxation toward the force-free state.
The coefficients ${c_h^2}$ and ${c_p^2}$ in Equation (\ref{eq:eq5}) are set to constant values of 0.04 and 0.1, respectively.

The initial condition was given by the potential field extrapolated from the observed $B_{z}$ using the Green function method \citep{Sakurai1982}. For the boundary conditions, the normal components of the magnetic field are fixed at all boundaries, while the tangential components follow the induction equation except at the bottom. Velocities are set to zero at all boundaries, and $\partial / \partial n = 0$ is imposed on $\phi$. At the bottom boundary, the tangential field is defined as $\bm{B}_{\text{bc}} = \gamma \bm{B}_{\text{obs}} + (1 - \gamma) \bm{B}_{\text{pot}}$, where $\bm{B}_{\text{bc}}$ corresponds to the tangential component, which represents a linear combination of the observed magnetic field ($\bm{B}_{\text{obs}}$) and the potential magnetic field ($\bm{B}_{\text{pot}}$). The parameter $\gamma$ ranges from 0 to 1. $\gamma$ is initialized to zero and increased by $d\gamma$ once the total Lorentz force, $R = \int |\bm{J}\times \bm{B}|^2 dV$, falls below $R_{\text{min}}$. We used $R_{\text{min}} = 5.0 \times 10^{-3}$ and $d\gamma = 0.02$. To suppress discontinuities near the boundary, the velocity is limited such that if the Alfvén Mach number $v^* = {|\bm{v}|}/{|\bm{v}{_\text{A}}|}$ exceeds $v{_\text{max}}$ (0.04), it is rescaled as $\bm{v} \rightarrow (v_{\text{max}}/v^*) \bm{v}$.
\subsection{Data-constrained MHD Simulation}\label{sec:mhd}
We carried out data-constrained MHD simulations using the NLFFF as the initial condition \citep{Inoue2014,Inoue2018} to study the evolution from flare initiation to eruption. The governing equations are the same as in the NLFFF calculation, but the bottom boundary treatment and the velocity limit differ. The normal magnetic field remains fixed in time, assuming stationary footpoints due to the slow photospheric response, while the tangential components evolve according to the induction equation with zero velocity. The resistivity and viscosity were set to $\eta = 1.0 \times 10^{-5}$ and $\nu = 1.0 \times 10^{-4}$. Both the NLFFF and MHD simulations used a computational domain of $362.5^3$ Mm$^3$ ($1.0^3$ in normalized units). The HMI data were resampled from $1000^2$ grids to $500^2$ grids by $2 \times 2$ binning. The simulation time is normalized by the Alfvén time, with $t=1.0$ corresponding to about six minutes.

\citet{Matsumoto2025b} have already clarified the MHD dynamics of the eruption for the X7.1 flare. The NLFFF revealed sheared field lines, between which tether-cutting reconnection occurred during the MHD simulation. These reconnected field lines formed a magnetic flux rope that subsequently erupted. After the reconnection, post-flare loops were formed in a configuration consistent with the AIA 171 \AA.

\section{Results} \label{sec:sec3}
\subsection{Hard X-ray Imaging observed by ASO-S/HXI}\label{sec:sec3.1}

\begin{figure*}[t]
\plotone{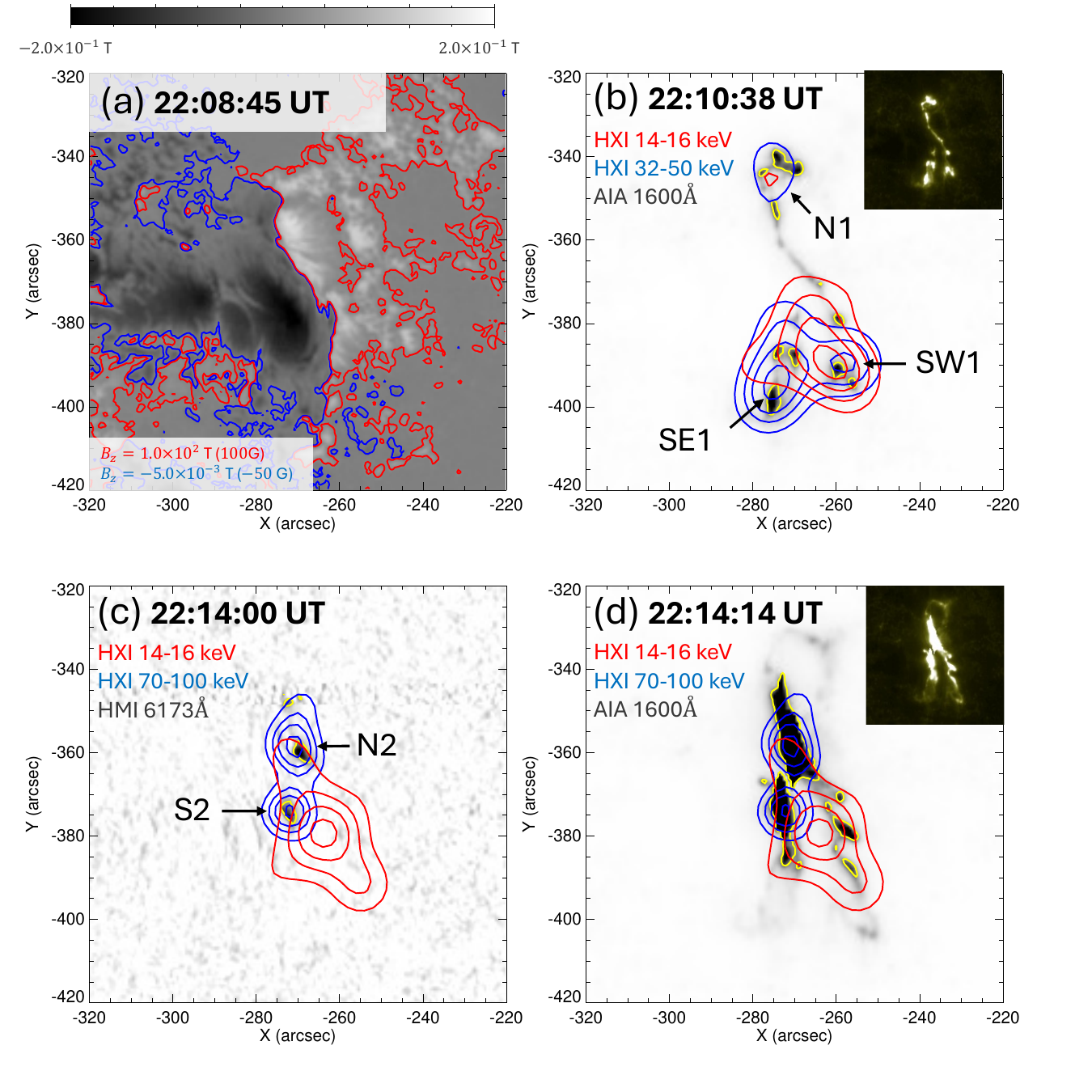}
\caption{(a) HMI magnetogram at 22:08:45 UT, with $B_z=1.0 \times 10^{-2}$ T in red and $B_z=-5.0 \times 10^{-3}$ T in blue. (b) AIA 1600 \AA\ at 22:10:38 UT with HXR sources observed by HXI (14-16 keV in red and 32--50 keV in blue at 22:10:30-22:10:50 UT). The blue contours correspond to 30\%, 50\%, 68\%, and 90\% of the peak, and red to 30\%, 50\%, 70\%, and 90\%. (c) HMI 6173 \AA\ difference image (22:14:00-22:12:30 UT) showing WL kernels in black at N2 and S2. Contours show HXI HXR sources (14-16 keV in red and 70--100 keV in blue). For HXR image synthesis, we selected the intervals 22:14:00-22:14:20 UT for the red contours and 22:13:40-22:14:40 UT for the blue contours. Contour levels are the same as in (b). (d) AIA 1600 \AA\ at 22:14:14 UT with the same contours as in (c). In panels (b)-(d), yellow contours show the enhanced region in each image.
}\label{fig:fig2}
\end{figure*}

 Figure \ref{fig:fig2}(a) shows the HMI magnetogram at 22:08:45 UT with $B_z=1.0 \times 10^{-2}$ T in red and $B_z=-5.0 \times 10^{-3}$ in blue. In Figure \ref{fig:fig2}(b), the AIA 1600 \AA\ image is overlaid with HXR sources observed by HXI in Phase 1. The southwestern footpoint source was observed during Phase 1 and is located in the positive polarity, hereafter referred to as SW1 (southwestern and Phase 1), while the southeastern one lies in the weak negative polarity (SE1), where $B_z$ is $\approx -5.0 \times 10^{-3}$ T, suggesting that they form conjugate footpoints of newly reconnected field lines associated with the main reconnection site. A faint northern source (N1) is visible. Whether its conjugate is SE1, SW1, or undetected is uncertain at this stage. For the imaging, the 32--50 keV range was used to capture non-thermal emission in Phase 1, while in Phase 2 (Figures \ref{fig:fig2}(c)–(d)), where the observation corresponds to a time closer to the HXR peak (see Figure \ref{fig:fig1}(a)), a higher energy band (70--100 keV) was selected to better capture the non-thermal component. In both phases, the 14--16 keV band was used to capture the thermal emission. We use the AIA UV ribbons as the reference during Phase 1 for co-alignment. In Phase 2, given the large magnitude of this X7.1 flare, we instead use the white light (WL) emission, which shows a strong spatio-temporal correlation with the HXR sources. To match UV ribbons in Phase 1, in Figure \ref{fig:fig2}(b), the HXI sources in the original image were shifted slightly by 3 arcsec westward. Figures \ref{fig:fig2}(c) and (d) show the spatial relationship between HXR, WL, and UV in Phase 2. In Phase 2, the northern and southern HXR footpoints are denoted as N2 and S2, respectively. For Figures \ref{fig:fig2}(c) and (d), the HXR sources were shifted by 5 arcsec westward and 2 arcsec southward, based on the position of more compact WL kernels. These adjustments in Phases 1 and 2 are both realistic and necessary (see also e.g., \citealt{Ryan2024a}), yielding good spatial agreement between the HXR footpoints, the WL kernel, and UV flare ribbons. From Phase 1 to Phase 2, the UV flare ribbon, the thermal HXR source, and the main HXR footpoint sources progress along the polarity inversion line (PIL), indicating a spatial migration of the main reconnection site in the corona and emphasizing the need for a 3D understanding of the flare evolution.

\subsection{Magnetic Field Lines Traced from the Hard X-ray Footpoint Sources}\label{sec:sec3.2}

\begin{figure*}[t]
\plotone{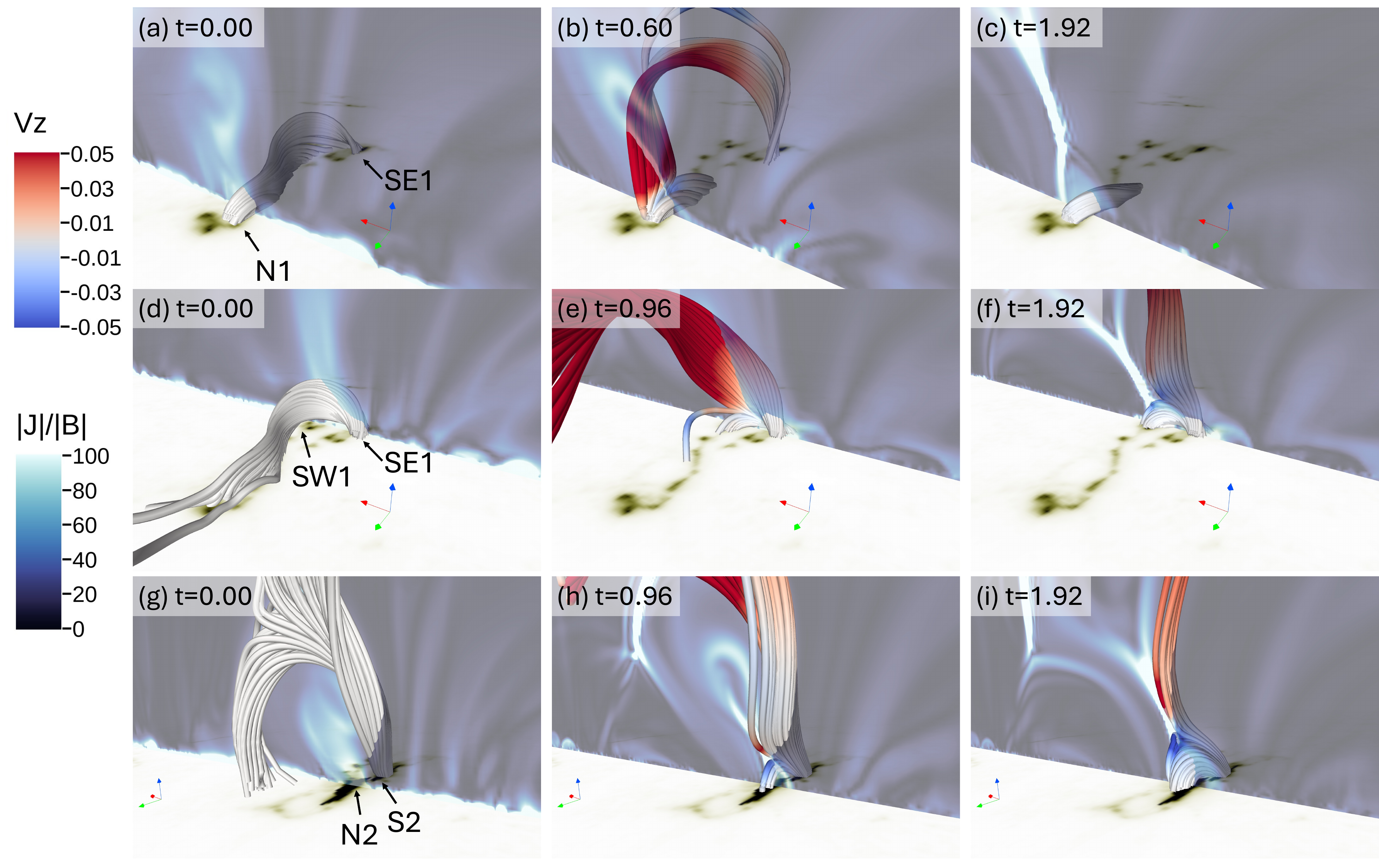}
\caption{Temporal evolution of the MHD simulation to show magnetic field lines traced from three locations (N1, SE1, and S2). All panels are presented from the northward viewing perspective. The top rows (a)-(c) show magnetic field lines traced from N1, the middle row (d)-(f) from SE1, and the bottom row (g)-(i) from S2. The field lines are colored by $V_{z}$, and the vertical cross-sections in each panel show $|{\bm J}|/|{\bm B}|$, allowing the current sheet structure to be identified. In the top and middle rows, the bottom image in each panel corresponds to the AIA 1600 \AA\ image in Phase 1 (Figure \ref{fig:fig2}(b)), while the bottom row corresponds to Phase 2 (Figure \ref{fig:fig2}(d)). Red, green, and blue arrows in each panel show the $x$, $y$, and $z$ axes in the simulation box. $t$ is the time normalized by Alfvén time in the simulation, as mentioned in Section \ref{sec:mhd}. An animation of the temporal evolution of this figure is available. The animation proceeds from t = 0.00 to 2.40 and shows magnetic field lines traced from N1 (top row), SE1 (middle row), and S2 (bottom row).
}\label{fig:fig3}
\end{figure*}

Figures \ref{fig:fig3}(a)-(c) show magnetic field lines traced from N1, the middle row (d)–(f) from SE1, and the bottom row (g)-(i) from S2. Field lines are colored by $V_{z}$, and the vertical cross-sections of $|{\bm J}|/|{\bm B}|$ highlight the current sheet. For Figure \ref{fig:fig3}, the AIA 1600 \AA\ images were reprojected from helioprojective coordinates into the CEA coordinate system of the HMI boundary of the MHD simulation, enabling a consistent overlay of the magnetic field lines with the flare ribbons and HXR footpoints via the AIA image as a reference. First, the field lines traced from N1 were initially long and twisted in the NLFFF, but they evolved into shorter loops through reconnection. Before this conversion, these field lines briefly join the erupting system and connect to SE1 in Figure \ref{fig:fig3}(b), so N1 and SE1 appear as conjugate HXR footpoints of erupting magnetic flux ropes, as reported in recent observations \citep{Chen2020ApJ,Stiefel2023,Chen2025}. Next, the field lines traced from SE1 show the reconnected loops and eventually connect to the SW1 footpoint. Moreover, the simulation also reveals reconnected loops connecting S2 and N2. Therefore, the MHD simulation reproduces the reconnection process and confirms the presence of observed conjugate footpoints in Phases 1 and 2, emphasizing the importance of employing realistic 3D magnetic fields.

\subsection{Hard X-ray Imaging observed by Solar Orbiter/STIX}\label{sec:sec3.3}

\begin{figure*}[t]
\plotone{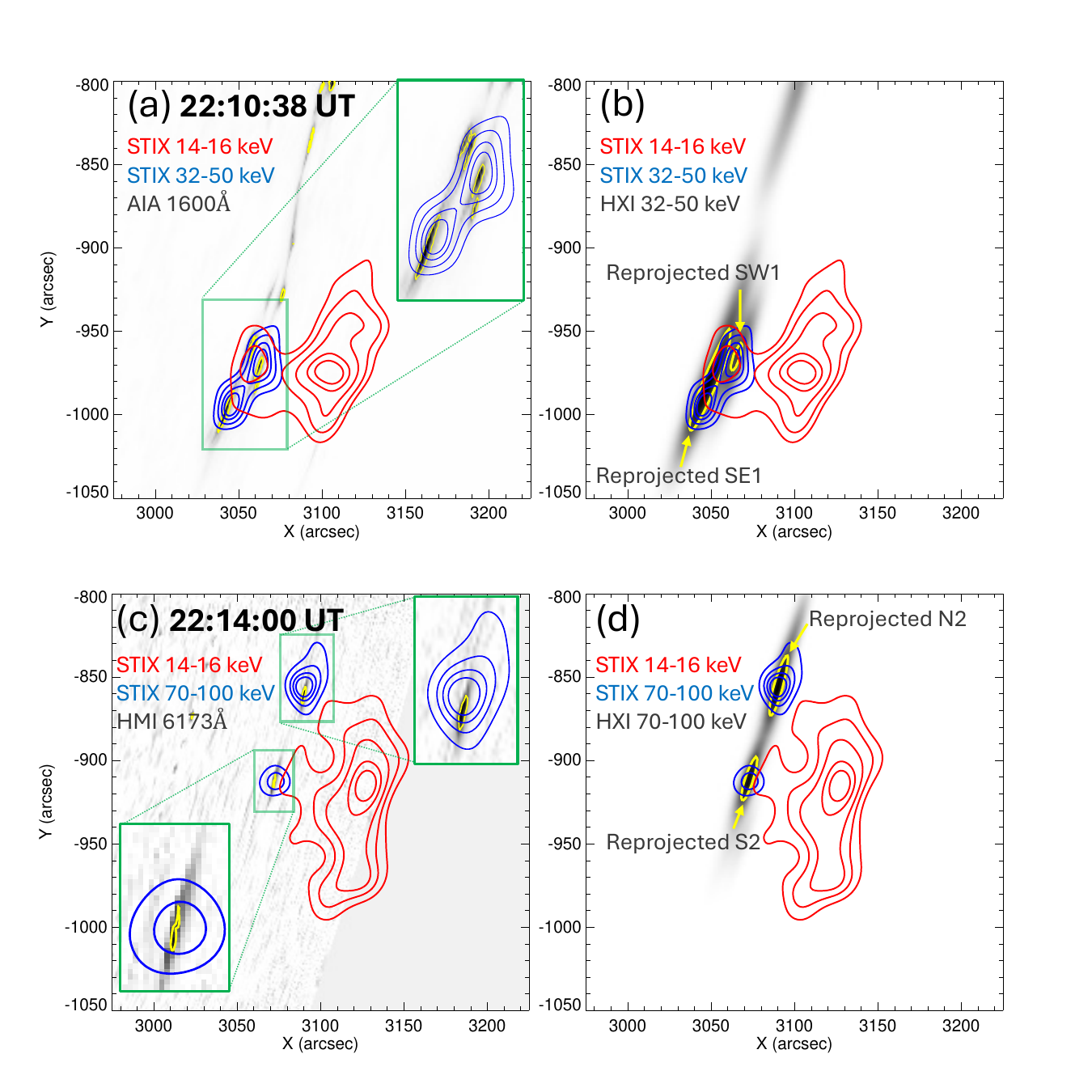}
\caption{(a)-(d) Reprojected AIA, HXR, and HMI images from Figure \ref{fig:fig2}(b) and (c) to the STIX viewpoint. (a) AIA 1600 \AA\ at 22:10:38 UT with HXR sources from STIX (14-16 keV in red, 32--50 keV in blue, 22:10:30-22:10:50 UT). Blue contours show 15\%, 30\%, 50\%, and 70\% of the peak intensity. Red contours show 11\%, 20\%, 35\%, 60\%, and 80\%. (b) HXI HXR in 32--50 keV as the background in Figure \ref{fig:fig2}(b) and the same STIX HXR contours in panel (a). (c) HMI 6173 \AA\ difference image (22:14:00-22:12:30 UT) to see WL kernels with STIX HXR contours (14-16 keV in red at 22:14:00-22:14:20 UT and 70--100 keV in blue at 22:13:40-22:14:40~UT). Contour levels are as in (a). (d) HXI HXR in 70--100 keV as the background in Figure \ref{fig:fig2} (c) and the same STIX HXR contours as in panel (c). In all panels, yellow contours show the enhanced region in each image.
}\label{fig:fig4}
\end{figure*}

Figures \ref{fig:fig4}(a)–(d) show the AIA, HXI, and HMI images from Figures \ref{fig:fig2}(b)-(c), reprojected to the STIX viewpoint. The contours in Figures \ref{fig:fig4}(a)–(d) represent the STIX HXR sources, with panels (a)–(b) corresponding to Phase 1 and panels (c)–(d) to Phase 2. To align the STIX HXR sources with the UV flare ribbons and HXI HXR footpoint source in Phase 1, a shift of 25 arcsec eastward and 25 arcsec northward was applied. Although this shifting of STIX appears large compared to that of HXI, it is justified by the pointing of STIX, as discussed by \citet{Massa2022}. For Phase 2, a 25 arcsec shift eastward and a 22 arcsec shift northward were applied to align the STIX HXR sources with the WL kernels. These shifts in Phases 1 and 2 result in good spatial agreement of the STIX HXR footpoint sources with the reprojected UV flare ribbons, WL kernels, and HXI HXR footpoint sources. In the STIX view, the thermal HXR emission exhibits a vertical extension. This feature is inaccessible to HXI alone, revealing thermal emission extending from the footpoints to the loop top and thus the vertical structure of the flare. Notably, during Phase 1, the thermal emission is intense near $y=-975$ arcsec, directly above the two footpoint sources, which coincide with the HXI footpoints SE1 and SW1 (Figure \ref{fig:fig2}(b) and reprojected in Figure \ref{fig:fig4}(b)). In contrast, during Phase 2, the emission centroid moved to $y=-910$ arcsec, above the STIX footpoints corresponding to N2 and S2 (Figures \ref{fig:fig2}(c)-(d) and Figure \ref{fig:fig4}(d)). This is consistent with an evolving reconnection site, whose apparent location changes over time as the flare progresses. The observed movement of the main reconnection site is confirmed by both HXI (Section \ref{sec:sec3.1}) and STIX observations.

\subsection{QSL Topology and Progression of Reconnection along the PIL}\label{sec:sec3.4}

\begin{figure*}[t]
\plotone{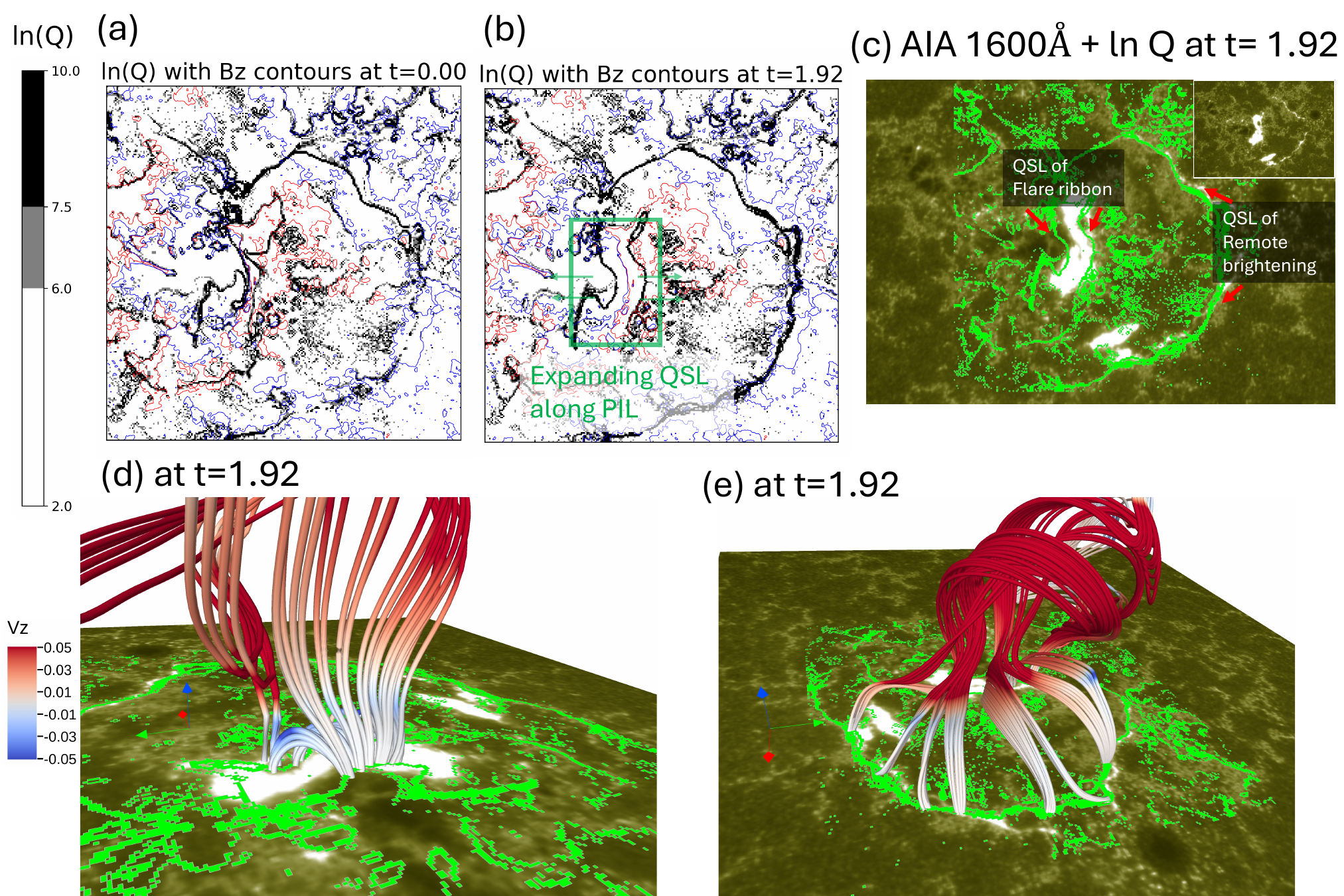}
\caption{(a) Logarithmic $Q$ at $t=0.00$. Red and blue contours mark $B_z=1.0\times10^{-2}$ T and $B_z=-5.0\times10^{-3}$ T at 20:36 UT. (b) Logarithmic $Q$ at $t=1.92$ with the same $B_z$ contours as in (a). (c) AIA 1600 \AA\ image at 22:10:38 UT with logarithmic $Q$ at $t=1.92$ overlaid, where regions with $\log Q \ge 6$ are highlighted in green for better visualization with the AIA image. The inset shows only the AIA image in the same field of view as (c). (d) Magnetic field lines at $t=1.92$ traced from QSL associated with the flare ribbons. (e) Magnetic field lines at $t=1.92$ traced from QSL associated with the remote brightening. Time $t$ is normalized to the Alfvén time (see Section \ref{sec:mhd}). An animation of the temporal evolution of panels (a) and (b) is available. The animation proceeds from t = 0.00 to 2.40.
}
\label{fig:fig5}
\end{figure*}

To assess whether the reconnection associated with SE1/SW1 (Phase 1) and N2/S2 (Phase 2) reflects two distinct sites or a single system evolving along the PIL, we calculated the squashing factor $Q$ for quasi-separatrix layers (QSLs). QSLs are regions where the connectivity of the magnetic field changes dynamically and are introduced to explain the 3D reconnection in no null point \citep{Priest1995}. The QSLs can be defined with high $Q \gg 2$ \citep{Titov2002}. Figures \ref{fig:fig5}(a) and (b) show $\log Q$ at $t=0.00$ and $t=1.92$, respectively. The continuous QSL footpoints run along the PIL and broaden with time, indicating that the Phase 1 and Phase 2 activity belongs to the same QSL system. Such a topology indicates slipping/ slip-running reconnection, in which footpoints drift along the QSLs as reconnection proceeds \citep{Aulanier2006, Aulanier2007, Lorincik2025}. Figure \ref{fig:fig5}(c) overlays the $\log Q$ map at $t=1.92$ on the AIA 1600 \AA\ image at 22:10:38 UT, showing correspondence between QSL footpoints and both the UV flare ribbons and the remote brightening. Field lines traced from QSL of the ribbons at $t=1.92$ in Figure \ref{fig:fig5}(d) demonstrate continuous reconnection along the PIL and link both HXR pairs (SE1–SW1 and N2–S2) within a single reconnecting system, rather than requiring two isolated sites. The temporal separation of the non-thermal HXR peaks (Phases 1 and 2) reflects episodic progression within one extended QSL system. Field lines traced from QSL regions at the remote brightening in Figure \ref{fig:fig5}(e) include erupting field lines, indicating interaction between the rising core field and overlying arcades, consistent with scenarios in which erupting core fields reconnect with ambient coronal loops \citep{Gibson2008}.

\section{Discussion} \label{sec:sec4}
\subsection{Two-Stage Evolution of Reconnection and Particle Acceleration}\label{sec:sec4.1}
We divided the impulsive phase into Phase 1 and Phase 2. The HXR footpoint sources occur at different locations in these intervals, indicating that the main reconnection site changed spatially, as discussed in Section \ref{sec:sec3.1}. As shown in the 32--50 keV light curve in Figure \ref{fig:fig1}(a), Phase 1 features a smaller HXR peak, whereas Phase 2 yields a larger one, suggesting the two peaks as evidence for two-stage particle acceleration \citep{Ning2018, Wang2025}. In Phase 1, observations are consistent with tether-cutting reconnection at lower altitudes. In Phase 2, in addition to the reconnection that developed in Phase 1, a more dominant reconnection occurs at higher altitudes, where the current sheet has further extended, leading to the enhanced HXR peak observed in Phase 2. This interpretation is further supported by Figures \ref{fig:fig3}(d) and (g). The field lines in Phase 1 are twisted and lie at lower altitudes, whereas those in Phase 2 are initially less twisted and located at higher altitudes. This is consistent with reconnection occurring higher in the corona during Phase 2. Quantitative acceleration measures, such as the evolution of the spectral index of the non-thermal electrons, require HXR spectroscopy and are left for future work. Our main findings focus on the HXR footpoint locations and their spatial agreement with the connectivity of the magnetic fields in the MHD simulation.

\subsection{Modeling Scope in this Study and Future Extensions}\label{sec:sec4.2}
Our analysis targets the spatiotemporal evolution of magnetic field lines and their registration with HXR sources. Within this scope, a zero-$\beta$ data-constrained MHD model is sufficient. We do not attempt to resolve microphysics (e.g., termination shocks or compressive heating), and omitting finite-$\beta$ effects does not alter our conclusions. Indeed, comparisons with and without plasma-density evolution show only minor quantitative differences while preserving topology and large-scale dynamics \citep{Inoue2014}. Looking ahead, finite-$\beta$ data-constrained simulations will be essential when the goals extend to modeling termination shocks, plasma heating, and compression and to constraining particle acceleration within a thermodynamically consistent framework. As a next step to better understand the particle acceleration and transport processes, it is desirable to forward-model HXR and microwave emission from finite-$\beta$ data-constrained 3D MHD snapshots, using tools such as \texttt{GX Simulator} \citep{Nita2023ApJS} that adopts a parametrized non-thermal electron distribution, or by performing self-consistent macroscopic particle simulations \citep{Chen2024, Li2025}, to produce synthetic images and spectra for quantitative comparison with observations.

A separate challenge is the timing mismatch between the simulation and the actual flare. In our data-constrained model, the horizontal magnetic fields at the bottom boundary evolve according to the induction equation. Although this setup successfully reproduces the eruption dynamics, it results in reconnection occurring earlier than observed. In reality, the slower evolution of photospheric motions likely constrains the magnetic fields, leading to a more gradual reconnection process. To achieve closer temporal correspondence, future studies could employ fully data-driven MHD simulations that are constrained by a time series of photospheric magnetic fields \citep{Jiang2016, Hayashi2018}.

\section{Summary} \label{sec:sec5}
We investigated the 3D magnetic configuration of the X7.1 flare that occurred on October 1, 2024, in NOAA AR 13842, using a combination of stereoscopic HXR observations from STIX and HXI, and data-constrained MHD simulations. The comparison between HXR observations and the magnetic field reproduced by the data-constrained MHD simulation demonstrates that such simulations can provide a realistic 3D magnetic environment for investigating non-thermal radiation signatures and particle acceleration processes in solar flares. 

From the HXI observations, we identified non-thermal footpoint sources in two phases (Phases 1 and 2). Between these phases, the main non-thermal footpoints progressed together with the thermal HXR emission and the UV ribbons. Tracing magnetic field lines from these footpoint locations in the MHD simulation revealed the reconnected loops whose footpoints match the observed conjugate HXR footpoints in Phases 1 and 2. These results provide direct support from the MHD modeling for the dynamics inferred from HXR observed by HXI. The stereoscopic perspective of STIX added a critical dimension by revealing the vertical extension of the thermal HXR emission. Due to projection effects, this is a feature inaccessible to HXI alone, showing thermal emission extending from the footpoints to the looptop. Moreover, the movement of both the thermal emission and the footpoint sources from Phase 1 to Phase 2 further corroborates the dynamics identified by HXI and AIA, strengthening the conclusion that the main reconnection site progressed during the flare. Together with the continuous and expanding footpoints of the QSL during the simulation, the observed migration between Phases 1 and 2 is interpreted as episodic energy release progressing along one extended reconnection system. Note that the model is designed to reproduce the 3D topology and connectivity of the magnetic field lines, while the absolute order of reconnection is not expected to match the observations one-to-one. These results emphasize the value of combining multi-perspective HXR observations with data-constrained MHD simulations to capture the 3D dynamics of solar flares. Although the zero-$\beta$ MHD framework does not capture termination shocks, which can be crucial for electron acceleration \citep{Tsuneta1998,Chen2015}, future work with finite-$\beta$ data-constrained MHD, macroscopic particle simulations, HXR/microwave spectral analysis, and emission modeling will more realistically constrain particle acceleration in observed magnetic fields.

\begin{acknowledgments}
This study is supported by NASA grants 80NSSC23K0406, 80NSSC21K1671, 80NSSC21K0003, 80NSCC24M0174, 80NSSC24K1242, and NSF grants AST-2204384,  AGS-2145253, 2149748, 2206424, 2309939 and 2401229. This work was carried out with the partial support of the ISEE International Joint Research Program at Nagoya University, Japan. The 3D visualizations were produced using VAPOR (\href{http://www.vapor.ucar.edu}{\texttt{www.vapor.ucar.edu}}), a product of the National Center for Atmospheric Research \citep{Li2019}. All calculations of the MHD simulations in this study were performed using the computing facilities of the High Performance Computing Center (HPCC) at the New Jersey Institute of Technology. The ASO-S mission is supported by the Strategic Priority Research Program on Space Science, the Chinese Academy of Sciences, Grant No. XDA15320000. Solar Orbiter is a mission of international cooperation between ESA and NASA, operated by ESA \citep{Muller2020}. The STIX instrument is an international collaboration between Switzerland, Poland, France, Czech Republic, Germany, Austria, Ireland, and Italy. SK is supported by Swiss PRODEX grant for STIX. The HXI and STIX datasets analyzed in this study are publicly available from the Data Archives (\url{http://aso-s.pmo.ac.cn/sodc/dataArchive.jsp} and \url{https://soar.esac.esa.int/soar/}).

\end{acknowledgments}

\bibliography{sample631}{}
\bibliographystyle{aasjournal}
%




\end{document}